\documentclass[12pt]{article}
\usepackage[super,sort&compress,comma]{natbib} 
\usepackage{mhchem}
\usepackage{sectsty}
\usepackage{balance} 
\usepackage{amssymb}
\usepackage{graphicx} 
\usepackage{lastpage}
\usepackage{fancyhdr}
\usepackage{color}
\date {}

\begin{document}
\renewcommand{\thefootnote}{\fnsymbol{footnote}}

\noindent\LARGE{\textbf{Gene silencing and large-scale domain structure of the E.~coli
  genome}}
\vspace{0.6cm}

\noindent\large{\textbf{Mina Zarei,\textit{$^{a}$}\textit{$^{b}$}\textit{$^{c}$} Bianca Sclavi,\textit{$^{d}$} and
Marco Cosentino Lagomarsino$^{\ast}$\textit{$^{a}$}\textit{$^{b}$}\textit{$^{e}$}\vspace{0.5cm}}}

\begin{abstract}
  The H-NS chromosome-organizing protein in \emph{E.~coli} can
  stabilize genomic DNA loops, and form oligomeric structures
  connected to repression of gene expression.
  Motivated by the link between chromosome organization, protein
  binding and gene expression, we analyzed publicly available genomic
  data sets of various origins, from genome-wide protein binding
  profiles to evolutionary information, exploring the connections
  between chromosomal organization, gene-silencing, pseudo-gene
  localization and horizontal gene transfer.
  We report the existence of transcriptionally silent contiguous areas
  corresponding to large regions of H-NS protein binding along the genome, their position indicates a possible relationship with the
known large-scale features of chromosome
  organization. 
\vspace{0.5cm}
 \end{abstract}

\footnotetext{\textit{$^{a}$~G\'enophysique / Genomic Physics Group, UMR~7238 CNRS
  "Microorganism Genomics", Paris, France.  E-mail: Marco.Cosentino-Lagomarsino@upmc.fr }}
\footnotetext{\textit{$^{b}$~University Pierre et Marie Curie, 15 rue de l'\'Ecole de
  M\'edecine, 75006, Paris, France. }}
\footnotetext{\textit{$^{c}$~Dipartimento di Fisica, Universit\`a degli Studi di Milano,
  via Celoria 16, 20133 Milano, Italy. }}
\footnotetext{\textit{$^{d}$~LBPA, UMR 8113 du CNRS, Ecole Normale Supérieure de Cachan,
  61 Avenue du Président Wilson, 94235 CACHAN, France. }}
\footnotetext{\textit{$^{e}$~Dipartimento di Fisica, Universit\`a di Torino, via
  P.~Giuria 1, Torino Italy.
 }}

\section{Introduction}

Bacterial chromosomes are several orders of magnitude longer than
their cells. Therefore, a chromosome must be tightly packed to fit
into the cell volume, while simultaneously being efficiently
transcribed and replicated. The molecular mechanisms responsible for the coordination of nucleoid
organization and gene expression remain to be  fully described~\cite{gino}. Bacterial DNA is condensed in a compact
DNA-protein complex called ``nucleoid'', largely dependent on a set of
shaping factors including the degree of supercoiling and the binding
of so-called ``nucleoid associated proteins'' (NAPs) such as Fis,
H-NS, IHF, HU and Dps~\cite{Dillon2010,Browning2010}.
The organization of the nucleoid presents ordered structures at
different scales: from DNA supercoiled loops of an approximate length
of 10 kb~\cite {Postow2004} to larger scale structures organizing the
circular DNA polymer in four macrodomains which range from 0.6 to 1 Mb
in length. Macrodomains were initially detected by analyzing the
recombination efficiency between pairs of sites scattered along the
chromosome and subsequently they were observed to spatially demix in
the cell.  In addition to macrodomains, two less structured regions
exist, the function and organization of which are not completely
understood~\cite{matp1,matp2,macrodomain,isolate}.  Some NAPs present
a macrodomain-specific distribution of binding sites,
suggesting their implication in the definition of macrodomains~\cite
{matp1,matp2}.
The physical organization of the bacterial chromosome can also affect
the accessibility and activity of large sets of genes, thus ultimately
regulating their expression, thereby contributing to the adaptation to
changes in environmental conditions and stresses~\cite{Dillon2010,Sobetzko2012,blot}.

There is thus a link between chromosome organization and gene
expression or silencing, which is possibly subject to natural
selection.
A particularly interesting NAP is the H-NS protein, which can
stabilize DNA loops created by supercoiling~\cite{Dillon2010,DLK+05}. This protein preferentially binds to curved
and AT-rich DNA~\cite{curve}, and is also reported to form oligomeric
nucleoprotein complexes that can act to form stable bridges between
distant regions of the chromosome~\cite{Dame2006,Luijsterburg2006}.
H-NS is generally believed to play a role as a global transcriptional
silencer. Its binding affects the transcription of specific genomic
contiguous areas under nucleoid perturbations affecting the global
levels of supercoiling~\cite{scolari,MGH+08}.  Intriguingly, this
protein is also implicated in selectively silencing the transcription
of horizontally acquired, AT-rich genes, including elements of
pathogenicity islands in
enterobacteria~\cite{Dorman2007a,oshima,hns-hgt}.
A recent high-resolution imaging study indicates that, in contrast
with other NAPs, H-NS can form compact subcellular foci bridging
together distant chromosomal loci. It was shown that this process is
dependent on oligomerization of DNA-bound H-NS and that deleting H-NS
leads to extensive chromosome restructuring~\cite{Wang2011a}.
In other words, there seems to be a connection between the mechanisms
that cells have evolved to efficiently integrate and regulate the
expression of newly acquired DNA into the genome and the physical
organization of the chromosome.

Given the complexity of the interplay and the feedback between
chromosome organization, gene expression regulation and evolution, a
statistical analysis at multiple scales is required to integrate data
coming from different high- and low-throughput experiments.
This work analyses multiple publicly available genomic data sets of
different origin, ranging from genome-wide protein binding profiles to
evolutionary information, exploring the connections between
chromosomal organization, gene-silencing, pseudo-gene localization and
the evolutionary dynamics of horizontal gene transfer.

We report the existence of transcriptionally silent contiguous areas
corresponding to regions of H-NS binding
along the genome. Notably, these areas are typically found near the
macrodomain boundaries and are enriched by pseudo-genes and
horizontally transferred genes, pointing to links between chromosome
organization and transcriptional silencing for functional and
evolutionary reasons. We also report intriguing changes in the size of
the contiguous areas bound by H-NS proximal to the terminus region as
cells progress through the exponential growth phase into stationary
phase.

\section{Materials and Methods}

\paragraph*{Reference genome. }  

$\ $The reference genome considered in this work was built as the union of
the sets of E.~coli K-12 annotated genes (including pseudo-genes and phantom genes) from the
RegulonDB~\cite{regulondb} and the EcoGene~\cite{ecogene}
databases. These gene set contains 4667 genes and includes all the
genes from smaller data sets that were used in our analyses.

\paragraph*{Pseudo-genes, prophages and horizontal  gene transfers
  data. } 

$\ $We used a list of 214 pseudo-genes retrieved from the
EcoGene~\cite{ecogene} database. In order to apply the linear
aggregation algorithm of Scolari {\it et al.}~\cite{scolari}, which
tests the clustering against randomized gene lists, we included the
pseudo-gene list in our reference genome. The list of prophages
``NRPS'' (non-redundant prophage set), contains 261 genes and was
retrieved from the prophage database~\cite{prophage}.
The list of horizontally transferred genes is retrieved from the
Horizontal Gene Transfer DataBase (HGT-DB)~\cite{hgt-db}.  In the
HGT-DB database, genes are considered to be putative horizontally
transferred genes if they have extraneous G+C content and codon usage,
they are over 300 bp long and they do not deviate from the average
amino-acid composition. We also used a list of horizontally
transferred genes from Lercher and Pal~\cite{lercher}, where transfer
events were detected based on phylogenetic gene- and species-tree
information, mapping gene gain and loss events onto a phylogenetic
species tree. This approach has the desired effect of ignoring
xenologous replacements of genes with identical functions and
properties.

\paragraph*{H-NS extended binding regions and extended protein
  occupancy data.}

$\ $We considered genome-wide H-NS binding data obtained by chromatin
immunoprecipitation combined with microarray (ChIP-chip) or sequencing
(ChIP-Seq) from refs.~\cite{grainger, oshima, aswin}. The main analyses
were performed on the ChIP-Seq data (E.~coli K-12 MG1655) from
Kahramanoglou {\it et al.}~\cite{aswin}, and the other sets were used
as comparisons. These experiments for the Kahramanoglou {\it et al.} dataset were performed at multiple
time-points of a growth curve in rich media. The data were then used
by the authors to detect the presence of putative H-NS binding sites, and to
define extended protein binding regions as statistically significant
aggregates of binding sites separated by less than 200bp.  We
considered the sets of genes located in extended H-NS binding regions
in different growth phases.

For comparison, the same procedure was applied to the data from Oshima
{\it et al.}~\cite{oshima}, obtained using a high-density
oligonucleotide chip (ChIP-chip analysis). This experiment was
performed with the W3110 strain of E.~coli K12 grown in LB medium. The
H-NS genome-wide binding profile was assessed on exponentially growing
cells. The W3110 strain is very similar to the MG1655 strain and in
the data analysis the genome coordinates of MG1655 were used. 

Extended protein occupancy domains (EPODs) were retrieved from Vora
\emph{at al}~\cite{epoddd}. Using a modified large-scale ChIP
analysis measuring generic protein occupancy on DNA, this
study~\cite{epoddd} identified extended ($>1$ Kb) protein occupancy
domains (EPODs) along the genome coordinate, which presumably play a
structuring role for the nucleoid. The authors classified EPODs into
two classes, highly expressed (heEPODs) and transcriptionally silent
(tsEPODs), using the median expression level across domains.  We
considered the set of annotated genes located in these domains. The
list of genes located in tsEPODs contains 241 genes, and the list of
genes located in heEPODs contains 280 genes.

\paragraph*{Large-scale deletion (LD) series.}

$\ $The information on large-scale non-lethal deletions was retrieved from
Hashimoto {\it et al.}~\cite{minimal}, where large-scale chromosomal
deletion mutants of E.~coli were constructed to identify a minimal
gene set.  These experiments include a first series of 163 MD
(medium-scale deletions) designed on the basis of a classification of
gene essentiality based on the available literature, followed by 75 MD
deletions using the E.~coli homologous recombination system. Based on
these results, Hashimoto \emph{et al.} constructed the long deletion units, termed the LD
(large-scale deletion) series.

\paragraph*{Macrodomains and chromosomal sectors.}

$\ $The location of the chromosomal macrodomains were obtained from
Valens \emph{et al.}~\cite{macrodomain} and considered together with
the chromosomal sectors defined by Mathelier and Carbone~\cite{sector}
from the correlation of codon bias indices with the level of gene expression. The exact
coordinates used here are presented in Supplementary Tables 1 and 2.

\paragraph*{Statistical analysis of spatial clusters.}

$\ $In order to score the aggregation of gene sets along the genome with a
statistical method, we used the algorithm of Scolari \emph{et
  al.}~\cite{scolari,nust}.  This method considers the density of genes at
different scales (bin-sizes) on the genome, and compares empirical
data with results from random shuffling null models. In order to avoid
spurious effects of binning, for each gene list a density histogram is
built by using a sliding window with a given bin-size. The resulting
plot of the averaged density of genes for every point of the circular
chromosome is considered at different observation scales of the
genome.
%
%
Density peaks with a significantly high number of genes are identified
by comparing empirical data with 10.000 realizations of a null
model. For every bin size, the null model considers the density
histogram from a random list of the same length of the empirical
one. The number of genes for every bin in the empirical histogram is
compared to the distribution of global maxima of the null model,
obtaining a P-value for the value of the empirical histogram for each
bin. This procedure enables the extraction of a list of statistically
significant ($P <0.01$) bin positions. For each bin-size (or
observation scale), clusters are defined as connected intervals
containing a significantly high proportion of the genes in the
list. 

\paragraph*{Sliding window histograms.} 

$\ $The sliding window histogram shows the density of a given gene set
along the genome for a given window size, or observation scale. We
used window sizes in the range 50-500 Kb.  Given a window size, the
number of genes falling inside the window is counted, while the window
slides along the genome stepwise with a step size of around 500bp (smaller
than the typical gene length).

\section{Results}

\subsection{Coherence of macrodomain structure with H-NS binding
  regions}


H-NS is a nucleoid associated protein and a global repressor of
transcription.  It is known to silence horizontally transferred
genes~\cite{hns-hgt,oshima}, and contributes to the formation of
transcriptionally silent regions~\cite{epoddd}. It is also known to
bind to specific DNA sequences and structures such as AT-rich and
intergenic regions and curved DNA~\cite{at-rich,curve}.
H-NS binding profiles were previously investigated on a genome-wide
scale using ChIP-chip and ChIP-Seq~\cite{grainger, oshima, aswin}, in
particular as a function of growth-phase in Kahramanoglou \emph{et
  al.}~\cite{aswin}.  The latter study identified extended protein
binding regions on the genome at different phases of the growth curve by joining
all statistically significant binding sites closer than 200 bp. The
detected binding signal was similar at all stages of growth.  However,
the number of H-NS binding regions and genes targeted for binding
increased as the cells progressed from exponential to stationary phase
(both due to stationary phase-specific binding regions and to
extension of mid-exponential phase binding regions).  They also found
that H-NS binds to significantly longer tracts of DNA than Fis, and
that differential gene expression upon deletion of H-NS is associated
with the length of a binding region, for example, the operons inside
long binding regions display increased differential expression,
indicating a greater degree of repression.

\begin{figure}[h]
\centering
    \includegraphics[width=110mm]{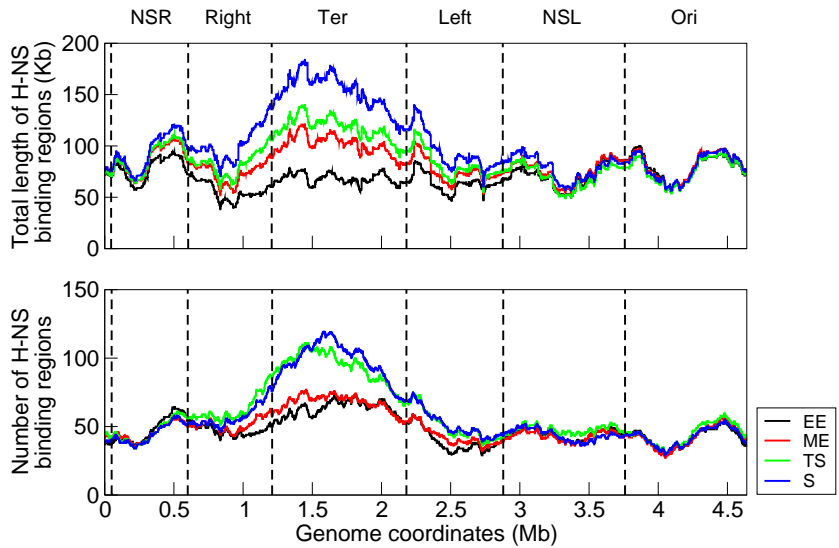}
    \caption{\label{growth}{\bf H-NS binding regions in the Ter macrodomain
        increase in length and number as cells progress from the early-exponential 
        to stationary growth phase.}  The plots show sliding
      window sums of the  length (top panel) and number (bottom panel) of H-NS binding regions along the
      genome in early-exponential (EE), mid-exponential (ME),
      transition to stationary (TS) and stationary (S) phases of
      growth (window size=500 Kb). The total length of H-NS binding
      regions in the approximate Ter macrodomain region increases continuously as
      the cells progress from exponential to stationary phase, while the number increases only between the ME and TS phases.}
 \end{figure}

We addressed the question of whether these H-NS binding regions are
equally long along the genome.  Figure~\ref{growth} shows a
sliding-window histogram of the length and the number of H-NS binding regions along
the genome (window-size=500 Kb). The number of binding regions increases between mid-exponential phase and "transition to stationary" phase. 
At the same time, as the cells
transition from exponential to stationary phase the total length of
H-NS binding regions increases continuously. 
This increase is specific to a large Ter-proximal
region (roughly associated with the Ter macrodomain) and is thus due to a combination of the addition of new regions and the growth of existing ones.  
The binding regions used in this analysis were obtained by comparing
the number of reads mapped to each binding region (normalized by the
total number of reads obtained for that sample) with the corresponding
value from the Mock-IP using a binomial test~\cite{aswin}.
Note that the Mock-IP data (which are used to correct for gene dosage and other
spurious effects) were available only for mid-exponential
phase. However, gene dosage should be more balanced along the genome
coordinate in stationary phase. In order to have an additional
control, we also considered data which were not corrected for the
Mock-IP (Supplementary Figure 1).  The qualitative pattern of the
total length and number of binding regions before and after the Mock-IP control does not change significantly in the Ter-proximal region. The change in this case concerns mainly a large
region around Ori in the early-exponential and mid-exponential data sets. We can thus exclude that the observed increase of length and number of H-NS binding regions around Ter is
  only an effect of gene dosage difference between the data and the Mock-IP. 

\begin{figure}[h]
    \centering
    \includegraphics[height=110mm]{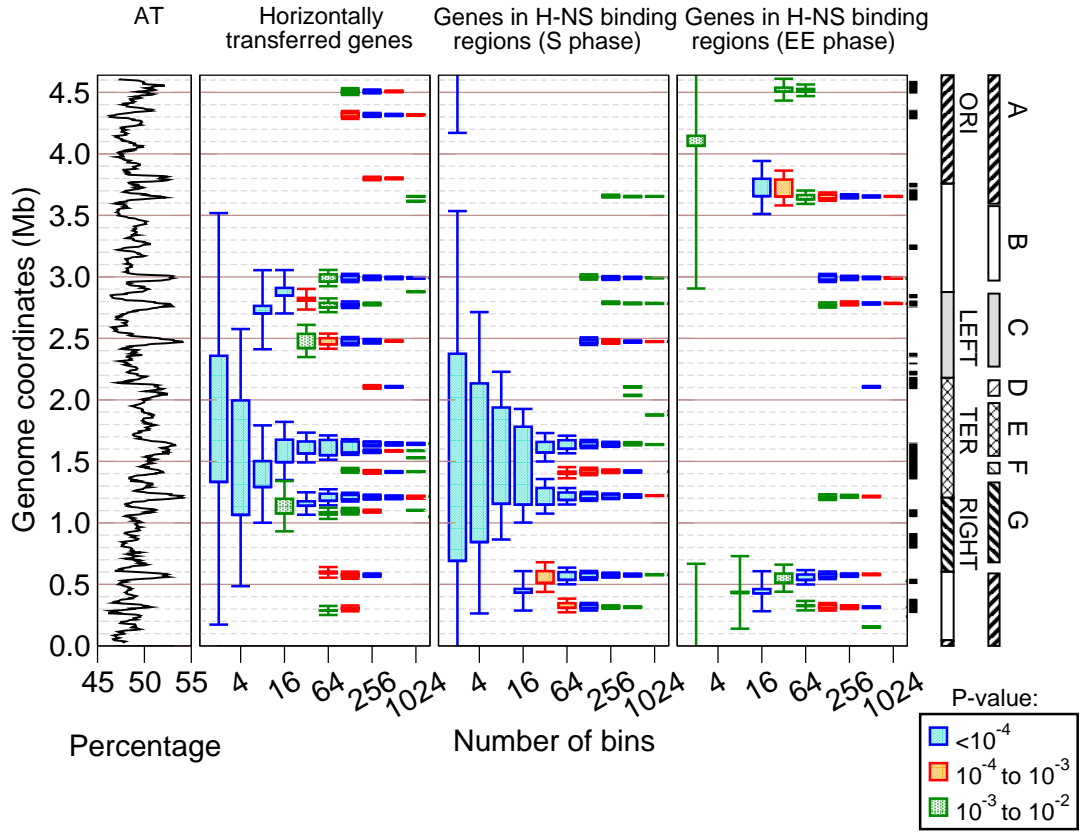}
    \caption{\label{hns-hgt} {\bf Genes located in the H-NS binding
        regions and horizontally transferred genes are clustered and
        often found close to  macrodomain boundaries.} The first panel shows sliding window averaging of
      the AT basepair percent along the genome coordinate (window size=50
      Kb). The second panel shows the diagram of the statistically significant clusters for horizontally transferred genes found using the algorithm
      presented by Scolari {\it et al.}~\cite{scolari}. The box indicates the position of
      the peak at a given scale of analysis (x axis) while the
      whiskers are the maximal extension of the cluster. The third panel shows the diagram of the
      statistically significant clusters for the genes located in the H-NS binding regions in stationary phase. The fourth panel
      shows a diagram of the statistically significant clusters found
      for the genes located in the H-NS binding regions in the early
      exponential growth phase~\cite{aswin}. The boxes on the
      right side of the plots show the large-scale deletions, (black rectangles), the macrodomains (labeled by their
      conventional names) and the sectors (labeled from A to G).}

    \end{figure}

Since H-NS binds to large regions along the genome, we wanted to
determine whether the genes located inside of these H-NS binding regions
were found in clusters along the genome.  We applied the linear
aggregation algorithm previously described~\cite{scolari}(see methods
section), for genes located in the H-NS binding regions found by
Kahramanoglou {\it et al.}~\cite{aswin}. Kahramanoglou {\it et al.}
identified H-NS binding regions for cells at different growth phases,
from early exponential to stationary phase.  The two panels on the
right of Figure~\ref{hns-hgt} and Supplementary Tables 3 and 4, show
the result of the linear aggregation analysis for genes located in the
H-NS binding regions in early exponential and stationary phase. One
can see that the genes located in the H-NS binding regions in early
exponential phase are clustered close to the macrodomain boundaries
and that this cluster pattern is preserved as the cells go from early
exponential to stationary phase (Figures~\ref{hns-hgt} and
Supplementary Figure 2). In stationary phase however, new clusters
appear inside the Ter macrodomain. We also performed linear
aggregation analysis on the H-NS binding targets obtained by Oshima {\it et
  al}~\cite{oshima}. The clusters found using data obtained by Oshima
{\it et al.} show similar clusters along the genome (Supplementary
Figure 3). While the macrodomain organization of the bacterial genome
was defined from experiments carried out in exponential
phase~\cite{macrodomain}, the large scale genome organization in
stationary phase remains to be defined. These results suggest that
while the boundaries of most macrodomains are maintained, the Ter
macrodomain becomes more silenced, and possibly more organized, in part by
H-NS binding.

\subsection{Horizontally acquired genes are clustered in the same
  genomic areas as H-NS binding regions}
The analysis presented so far points to coherent discrete regions
along the genome which are covered by H-NS binding regions and correlate with the macrodomain boundaries.
This suggests a possible link between a
structural, nucleoid-shaping role of these regions, and a functional
role of transcriptional silencing.

It is well known that the mechanisms that silence horizontally
acquired genes include the binding of H-NS~\cite{hns-hgt,oshima}. This
is due in part to the fact that transferred genes tend to be AT rich
and that H-NS has a higher affinity for this kind of DNA. We thus
searched for a relation between the position of the H-NS clusters and
the genes that are acquired by horizontal transfer along the genome.

To this end, we performed the linear aggregation analysis on the list
of horizontally transferred genes available from
HGT-DB~\cite{hgt-db}. Several methods have been used to detect
horizontally transferred genes based either on nucleotide composition
or on the failure to find a similar gene in closely related
species~\cite{rev-hgt}. The genes in the HGT-DB are found by the
former method. The results of our analysis show that horizontally transferred genes
are significantly clustered along the E.~coli genome (see the second
panel of Figure~\ref{hns-hgt} and Supplementary Table 5). The left panel of Figure~\ref{hns-hgt}
shows a sliding window histogram of the AT percentage along the
genome. In agreement with the method used for the HGT-DB one can see
that AT content along the genome correlates with the horizontal gene transfer (HGT)
clusters~\cite{at-rich}. As expected, the HGT clusters also have
similar localization with the clusters for the genes located in H-NS
binding regions.  In addition, some of these clusters also correspond to the location of
non-lethal large-scale deletions~\cite{minimal}.

\begin{figure}[h]
\centering
    \includegraphics[width=110mm]{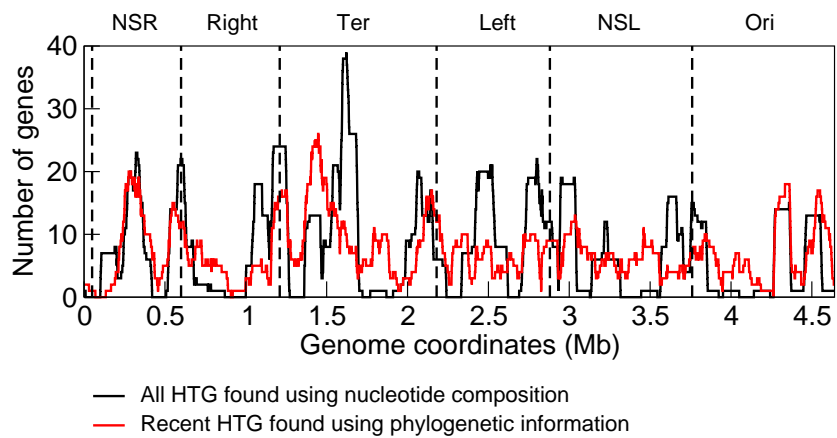}
    \caption{\label{lercher} {\bf Recently transferred genes are
        clustered along the genome.}  Comparison of the histograms (window size 100 Kb) 
      of horizontally transferred genes found using nucleotide
      composition~\cite{hgt-db} with those found from phylogenetic information to occur before 
      divergence between E.~coli and Salmonella~\cite{lercher}.}
    \end{figure}

The most accurate methods of HGT identification rely on phylogenetic
tree information because it allows one to estimate the relative date of the
transfers. The study of Lercher and Pal~\cite{lercher} defined a list of
horizontally transferred genes according to the number of branches
that separate E.~coli K-12 from the node of the tree where the transfer
is detected.  We divided the list of transferred genes into two
groups, "recent" and "old" transfers. In order to produce a sufficiently
large data set, we defined horizontally transferred genes with an age
less than 5 branches as the recent transfers. With this choice, "recent" transfers correspond to events 
that occurred before the divergence between E.~coli and Salmonella (about 100MY)~\cite{lercher}. The results of this
analysis show a correlation between the distribution of recently
transferred genes and the horizontally transferred genes from the
HGT-DB data set, Figure~\ref{lercher}.
No significant clusters are detected when the whole list of transferred genes
defined by the Lercher and Pal phylogenetic analysis is taken
independently of the date of transfer.

\subsection{Genes located in the transcriptionally silent extended
  protein occupancy domains (tsEPODs) are clustered in discrete
  genomic regions}

\begin{figure}[h]
    \centering
    \includegraphics[height=110mm]{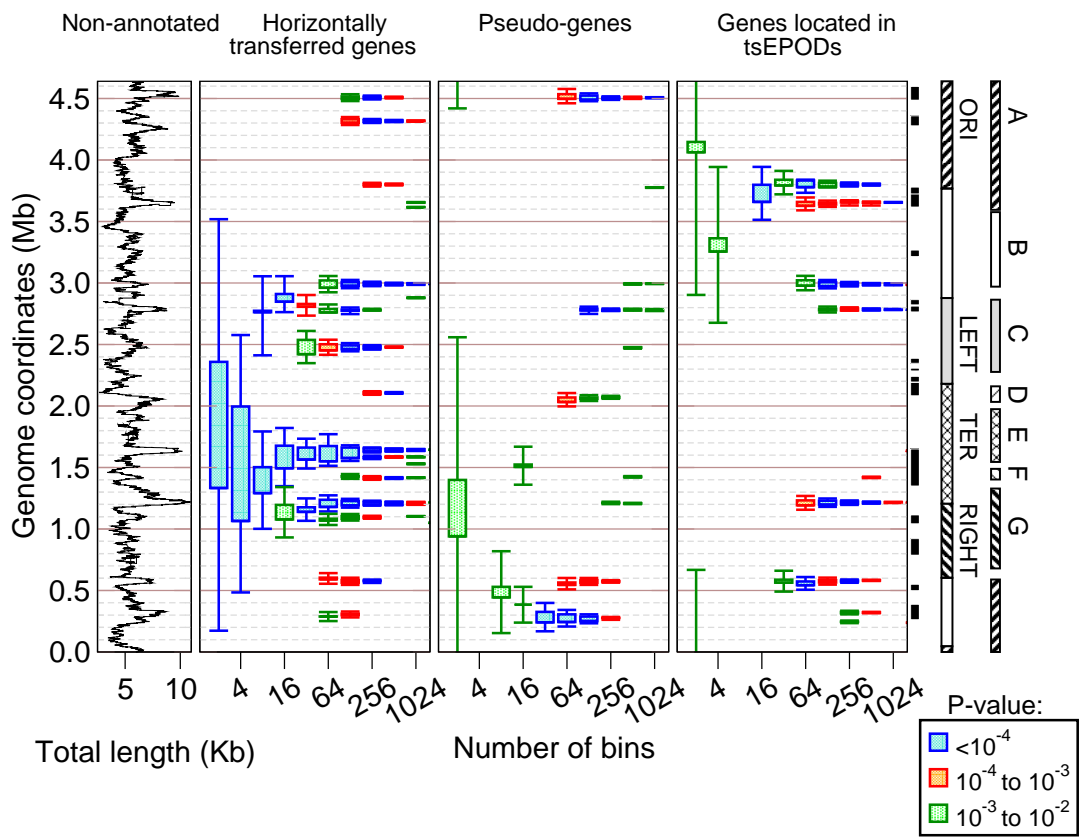}
    \caption{\label{silent} {\bf Gene sets overlapping with tsEPODs
        and pseudo-genes show significant and coherent linear
        aggregation clusters.}  The first panel shows sliding window histogram for total lengths of intergenic regions along the genome
      (window-size$\approx 36$ Kb). Here we considered pseudo-genes as the genomic regions. The second panel shows plot of the statistically significant clusters of horizontally transferred genes found using the algorithm
      presented by Scolari {\it et al.}~\cite{scolari}. The box indicates the
      position of the peak at a given scale of analysis (x axis) while
      the whiskers are the maximal extension of the cluster. The third panel shows a diagram of the
      statistically significant clusters of pseudo-genes. The fourth panel shows the diagram of the
      statistically significant clusters of tsEPODs. The boxes on the right side of
      the plots show the positions of the large-scale deletions (black rectangles), the macrodomains (labeled by their
      conventional names) and the sectors (labeled from A to G).}
 \end{figure}

It has been previously found that long H-NS binding regions overlap
with tsEPODs and are enriched with horizontally transferred
genes~\cite{aswin}.  We asked whether there are regions where the
genes that overlap with the tsEPOD domains are significantly clustered
along the genome coordinate. This analysis revealed multiple
significant clusters, as shown in the fourth panel of
Figure~\ref{silent} and Supplementary Figure 4. Notably, genes
overlapping with heEPODs and tsEPODs are clustered in complementary
genomic regions. The clusters of genes overlapping with heEPODs
include ribosomal and flagella genes (Supplementary Table
6). Conversely, tsEPODs are known to correspond to binding sites of
nucleoid-related proteins, and specifically of those of the H-NS
protein~\cite{epoddd, oshima}, and to areas enriched in hypothetical
ORFs. These observations are confirmed by the inspection of the tsEPODs
gene clusters.

The results shown in Figure~\ref{silent} show that most of the
clusters of genes located in tsEPODs are often correlated with the
boundaries of macrodomains~\cite{macrodomain} and are frequently part
of large-scale deletions that do not affect the
phenotype~\cite{minimal}.
Significant clusters are found near most macrodomain boundaries with
the exception of the Ter-Left and Ori-NSR macrodomain boundaries,
where a small group of tsEPODs gene clusters are found but not scored
as being significant by the algorithm.  Inside the Ter macrodomain we also
found a weakly significant clustered region correlated with the
boundary of the chromosomal sector E defined by Mathelier and
Carbone~\cite{sector}.
The only significant cluster of tsEPODs genes that does not
lie close to a macrodomain or chromosomal sector boundary is found at 0.245 Mb and contains CP4-6 prophage-like
genes~\cite{prophage}. 
\subsection{Pseudo-genes are clustered along the genome, and follow
  similar aggregation patterns as tsEPOD genes}

The positions and density of pseudo-genes were also included in the
analysis. Pseudo-genes have previously been reported to form from the inactivation of
 horizontally transferred genes~\cite{failed-hgt}.  Having established that
tsEPODs clusters appear to associate with macrodomain boundaries, we
proceeded to evaluate if similar global patterns could be reported for
pseudo-genes in general, which are, by definition, transcriptionally
silent.

The pseudo-gene list that we retrieved from the EcoGene
database~\cite{ecogene} contains 214 pseudo-genes. The list also
includes interrupted genes of the genome. The second panel of
Figure~\ref{silent} and Supplementary Table 7 show that pseudo-genes are clustered along the
genome and that the positions of these clusters often corresponds well with the
aggregation found for tsEPODs genes and with macrodomain and/or sector
chromosomal structure~\cite{macrodomain,sector}.
Note that the two gene sets have a significant intersection but are
naturally quite different from each other, as tsEPODs contain more
genes (Supplementary Tables 8 and 9).

Two significant clusters are found away from macrodomain boundaries,
the first, located around 0.27 Mb, overlaps with a corresponding
cluster of tsEPOD CP4-6 prophage-like
genes mentioned above, and with a large-scale
deletion with no phenotype~\cite{minimal}.
The second one (around 4.5 Mb) also corresponds to a large-scale
non-lethal deletion. We speculate that this could be compatible with
the Ori macrodomain limit on the right replichore, since intermediate coordinates were not probed by Valens {\it et al.}~\cite{macrodomain}.
The third panel of Figure~\ref{silent} shows the linear aggregation of
horizontally transferred genes as plotted in Figure~\ref{hns-hgt}.
One can see that all of the clusters from these different datasets correlate with each other.

\subsection{A non-uniform length distribution of non-annotated genomic
  regions points to additional links between gene silencing and
  chromosome domain structure.}

The data set considered in our analysis as the reference genome
includes genes, pseudo-genes and phantom genes (regions of DNA that previously identified as genes, 
which are not currently thought to be functional genes). Despite this, the  
regions between two consecutive elements (genes, phantom genes, or pseudo-genes) might
still contain silenced elements that are not detected. In order to
quantify the amount of DNA that is not annotated in one of the above
categories, we measured the inter-element distances and we calculated
their total length along the genome. The left panel of
Figure~\ref{silent} shows the sliding-window histogram of total
lengths of these non-annotated regions along the genome, with window
size $50$ Kb. Near the HGT clusters, and close to macrodomain
boundaries, the average length of the non-annotated sequences is
longer. Therefore, the clusters appear to contain less genetic information.

\subsection{Coherent aggregation of H-NS binding regions and silent
  genes with known chromosomal compartments }

\begin{figure}[h]
    \centering
    \includegraphics[height=100mm]{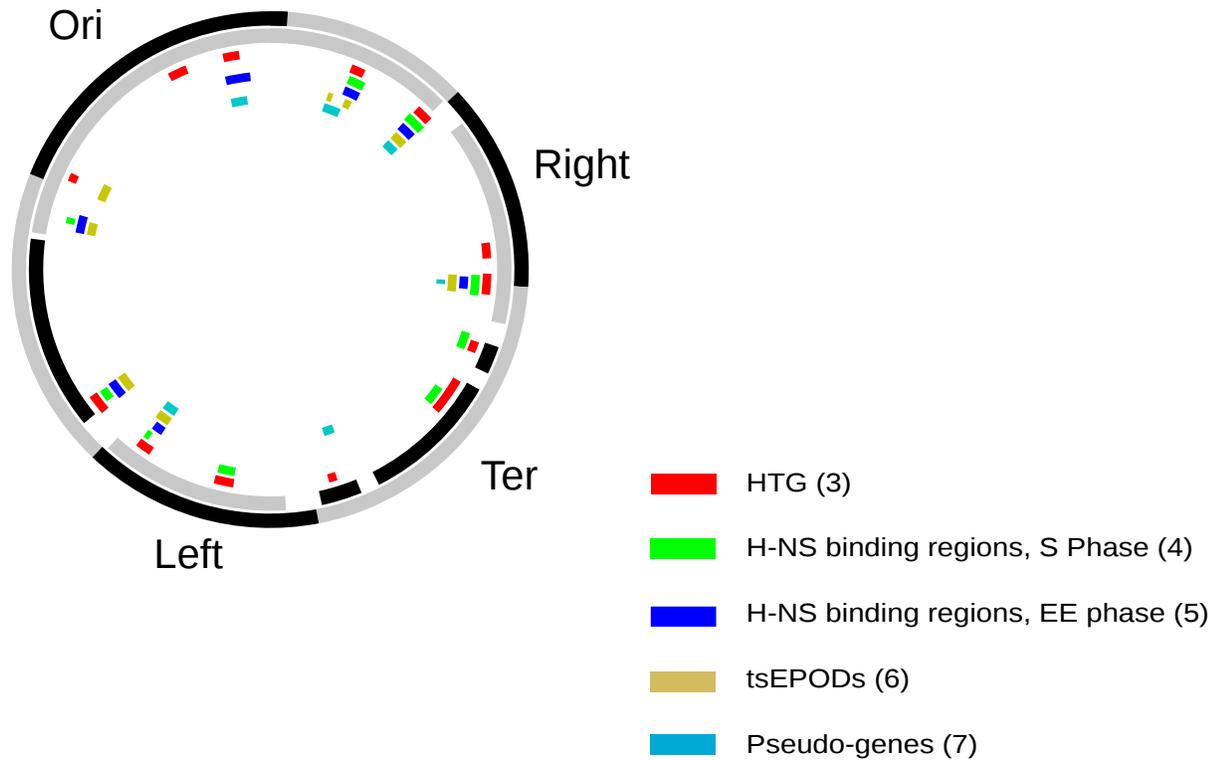}
    \caption{\label{circle}{\bf Coherence of the significant clusters
        found from different data sets.} Summary of the
      clusters found analysing different data sets. The outer circle
      shows the macrodomains~\cite{macrodomain}. The second
      circle shows the chromosomal sectors as defined by Mathelier and
      Carbone~\cite{sector}.  The remaining circles (third to seventh
      from the outer one) show the clusters found in
      horzontally transferred genes, genes located in H-NS binding
      regions in stationary and early exponential phases, genes
      located in tsEPODs, and pseudo-genes respectively. The diagram includes the
      clusters appearing when using bin sizes between 64 and 256 (preferentially reporting the coordinates at 128 bin).  }
 \end{figure}

Table~\ref{result} compares the significant clusters found using different data sets.  The listed regions were obtained by merging all
the overlapping clusters found using different datasets at
window-size$\approx$ 36 kb (bin-size=128). If a cluster is found in a
specific dataset it is labeled by a tick mark. Supplementary Table 10
shows the overlap between significant clusters found using different
data sets and large non-lethal deletions or prophages.  We can see
that many of the clusters overlap with prophages and non-lethal
deletions. Figure~\ref{circle} compares the significant
clusters found using the different data sets.  These clusters were
found by analyzing different datasets with a window-size of $\approx$ 36 kb
(bin size = 128). Figure~\ref{circle} also includes clusters detected at bins 64 or 256, but not 128. For example, the cluster found in the boundary between Right and Ter macrodomains using pseudo-genes data set refers to 256 bin.

Table~\ref{regions} summarizes the comparison between cluster
coordinates and the positions of macrodomain or sector boundaries, showing
the shortest distance between one edge of a cluster to the closest
macrodomain or chromosomal sector boundary in Mb, and the normalized
distance, which we defined as the same quantity rescaled by the length
of a macrodomain or sector where the cluster is found.  The position
of the clusters often correlates well with the closest macrodomain
boundaries. In order to verify whether there is a significant correlation between
these clusters and the boundaries of the macrodomains, we considered
randomly distributed non-overlapping clusters with the same number
(14) and sizes of the clusters reported in Table~\ref{regions}.
Comparing the average distance between the boundaries of the random clusters
and the closest macrodomains with the same measurement for the
empirical clusters, we found a P-value of 0.03. 
%
%

Supplementary Tables 8 and 9 show the intersection between different
gene sets used in this study, and the result of a hypergeometric test
assessing the statistical significance of the intersection.  These
data show that while the overlap of the gene sets is large, the sets
are far from being equal, and thus the observed spatial coherence in
their aggregation is \emph{a priori} not trivial.

We carried out systematic hypergeometric testing of enrichment for
MultiFun~\cite{MultiFun} functional categories in the lists of
functional genes located in each cluster and in the list of genes
located in all clusters. In the MultiFun data set 3382 out of 4667
genes have at least one annotation; within the clusters 757 out of
1075 genes have a MultiFun annotation. The results are reported in
Table~\ref{ontology1} and Supplementary Table 11, which summarize the
annotations with P-value $<0.01$.  Table~\ref{ontology1} also reports
the number of distinct operons in the list, useful in order to avoid
considering lists composed of only one or two operons as
significant. The genes located in the clusters appear to be involved
in membrane synthesis, pilus synthesis, anaerobic respiration,
lipopolysaccharide synthesis, phosphorous metabolism, phage related functions,
and protein folding. 
%
We did not observe a significant difference when the clusters’ function was
correlated with its distance from the macrodomains boundaries (Supplementary Table 11). Not surprisingly, most of the
clusters are enriched by prophage related functions. The clusters with
coordinates 3.6-4.3 Mb show enrichment in the terms membrane, surface
antigens, lipopolysaccharide, carbon compounds, and anaerobic
respiration.

Finally we wanted to test whether gene expression levels within the
clusters were significantly different from the average of the E.~coli transcriptome. To this end, we calculated gene expression levels within the clusters reported in Table~\ref{result}, using microarray data for exponentially growing cells~\cite{asap}
(Supplementary Figure 5). We found that most clusters have lower mean
expression than the genomic average, and that some of the clusters overlapping
with macrodomain boundaries have slightly higher mean
expression. We examined the genes responsible for this higher expression levels (Supplementary Figure 5). 
The cluster with coordinates 3.765-3.831 Mb is
expressed at a higher level than the average expression level of the
genome. There are two genes coding for ribosomal proteins rpmB and
rpmG (located at 3.81 Mb) inside of this cluster, which increase the
average expression of the cluster. In the cluster with coordinates
0.537-0.609 Mb there are two genes coding for outer membrane proteins,
ompt and nmpC, which are highly transcribed according to the E.~coli K-12 expression data
 we used (note however that nmpC has been previously reported to be silent within
E.~coli K-12~\cite{nmpc}).

\section{Discussion}


In this work we have shown that horizontally transferred genes, pseudo-genes, genes
in tsEPODs and H-NS binding regions are clustered in common regions
along the genome that often lay close to - or overlap with - macrodomain
boundaries or chromosomal sectors as defined by changes in
recombination rates, cellular location and codon usage~\cite{macrodomain,sector}.
We found that the transcriptionally silent Extended Protein Occupancy
Domains (tsEPODs) and the density of H-NS binding along the
genome~\cite{aswin,epoddd} show previously uncharacterized properties
linking genome evolution and organization. These areas form linear
clusters along the genome coordinate that are enriched in pseudo-genes
and horizontally transferred genes and also contain a higher proportion
of intergenic sequences. It was previously known that H-NS mediates
the silencing of horizontally acquired genes in
bacteria~\cite{hns-hgt} and that these H-NS binding regions correlate
with tsEPODS ~\cite{aswin}, meaning that presumably most of these
extended silenced region are under the regulatory control of H-NS
nucleoprotein complexes. Horizontally transferred sequences are more
AT-rich than the rest of the genome. H-NS has been shown to tightly
bind to AT-rich sequences, possibly because of their high intrinsic
curvature~\cite{at-rich,curve}.
The positions and density of pseudo-genes were also included in the
analysis. Pseudo-genes have already been reported to be connected to horizontally transferred genes, 
since a large fraction of prokaryotic pseudo-genes arise from partial genes acquired by
horizontal transfer events~\cite{failed-hgt}.

A more surprising fact is that often these clusters appear to be
coherent with the boundaries of macrodomains, and thus to be related
to the physical organization of the nucleoid.  Two different but non
exclusive mechanisms can account for the physical origin of
macrodomains. In the first, some organizing factors can bind to DNA and
create boundaries that physically separate the macrodomains. In the
second, the presence of specific determinants within the macrodomains
can cause their organization~\cite{gino}. According to Valens and
coworkers, the second scenario is more likely to apply to the E.~coli
chromosome~\cite{macrodomain}. This is supported by the
macrodomain-specific distribution of binding of the SeqA, SlmA and
MatP proteins found by genomics methods. It has also been shown
experimentally that the structuring of the Ter macrodomain relies on
the interaction of the MatP protein with a 13-bp motif called matS
found in multiple copies along the Ter macrodomain. Analogously  SlmA
and SeqA bind all along the genome, except for the Ter
macrodomain~\cite{matp1,matp2}. By contrast,  an additional
site-specific system appears to isolate the Ter macrodomain from other
parts of the chromosome. This system is constituted by two similar
12-bp sequences flanking the Ter macrodomain and one protein that is
required to isolate the Ter macrodomain~\cite{isolate}.
The results obtained in this work support complementary role of
the first mechanism of nucleoid organization. The positions of the
clusters identified here show extensive overlap with most macrodomain
boundaries, thus suggesting that they may contribute to the definition
of the macrodomains by acting as a kind of insulators to limit the local structuring of the genome. We can thus propose that silencing of gene
expression by H-NS is not the only result of the formation of these
clusters.  Evidence in support of this proposal comes from a study
aiming at defining a minimal genome. In this study it has been shown
that large scale deletions of non-essential genes result in important
changes in both nucleoid structure and in the morphological properties
of the cell, such as cell length and width~\cite{minimal}.  Not
surprisingly, we found that most of these large-scale non-lethal
deletions correlate well with the clusters found here, basically
composed of pseudo-genes and horizontally transferred genes. Thus deleting the
clusters could cause changes in macrodomain organization. We are
currently testing this hypothesis which is compatible with Wang {\it et al.} result on the H-NS mutants.

We now turn to the localization of horizontally transferred genes.  According to
the current picture, there should be no or little negative selection
against an insertion within a transcriptionally silenced region. In
addition, occasionally the insertion of the novel sequence may also
bring an advantage by allowing for the development a novel function to
the cell in the form of a new protein~\cite{selfish}.  The improved
structural organization of the nucleoid due to the presence of these
H-NS bound clusters may bring an additional advantage that may contribute to
their positive selection.
Finally, one may speculate that these prokaryotic clusters might play
 an analogous role as (constitutive) heterocromatin
in eukaryotes~\cite{human}. 

Together, this evidence suggests that the evolutionary fate of
horizontally transferred genes is initially bound to specific discrete
structural regions of the nucleoid, which are silenced by a
nucleoprotein complex and might also become ``spacers'' for
macrodomains.  
This might also explain the results obtained by the analysis of
recently transferred genes. It has been shown that transferred genes become more
GC-rich as they become integrated with the cellular regulatory and
protein-protein interaction network~\cite{lercher}. Therefore as a function of time
these genes would "move out" of the AT-rich and silenced clusters.

The analysis of the expression level of the genes in these clusters
shows that most of them are expressed at a level which is
significantly lower than the average for the E.~coli transcriptome
(Supplementary Figure 5). The analysis of the functional annotations
of the genes in the clusters indicates an enrichment for genes
involved in membrane synthesis, pilus synthesis, anaerobic respiration,
lipopolysaccharide synthesis, phosphorous metabolism and phage related functions.
These are genes whose products are located in cytoplasm, periplasmic space
and inner membrane.  In some cases, this enrichment is due to the
presence of large operons such as the lipid A-core biosynthesis operon in the
cluster between the Ori and NSL macrodomains. It may be possible that
the genes in these clusters are part of the genes induced in response
to membrane stress.
Finally, the enrichment of the clusters (reported in Table~\ref{result}) for
membrane proteins could in principle cause statistically increased
anchoring of the clusters to the membrane through the coupling of
transcription-translation and thereby affect nucloid organisation~\cite{membrane}, but this still remains highly speculative.


A second relevant result of this study is that the total length and number of
H-NS binding regions increases as the cells progress through
exponential phase into stationary phase and this happens specifically in
a large area close to the replication terminus, roughly compatible
with the Ter macrodomain.
The H-NS protein in the cell can be found in different states,
specifically bound to the DNA in either dimer or oligomer form,
non-specifically bound to the DNA and not-bound
~\cite{at-rich,Luijsterburg2006}.  The mechanism leading to large
H-NS-bound regions can be rationalized as  a nucleation-polymerization process,
where nucleoprotein oligomers grow progressively from specifically
bound H-NS nuclei~\cite{aswin,12200432}. This process would be
reminiscent of the case of the RecA DNA-repair protein~\cite{Bar-Ziv2002}. 

The first question to address is what can be the cause of the
specificity of this process to the Ter region. Possibly the simplest
explanation for this is that this region is generally more
AT-rich~\cite{at-rich}, which could increase the non-specific binding
affinity of H-NS for DNA, and thus locally speed up the
oligomerization dynamics, which is set by the one-dimensional
diffusion process of non-specifically bound proteins along the DNA.

A second relevant, and possibly more puzzling, question is why an
increase in H-NS oligomerization is observed at the terminus region in
stationary phase.  Let us suppose initially that the cellular
concentration of H-NS remains constant over the growth
curve~\cite{Luijsterburg2006}. One needs then to determine what is the
possible imbalance in chemical equilibria or kinetic factors between
this growth stage and the earlier ones. We can speculate that two
different phenomena can contribute to the growth of the H-NS binding
regions, the first is due to the change in the amount of DNA per cell
and the second is the decrease in DNA replication frequency as the
cells enter stationary phase.  The first phenomenon creates changes in
chemical equilibria, while the second could affect the kinetics of protein
oligomerization.

It is well known that, because of overlapping replication rounds, in fast growth
conditions the cells have a higher DNA content than in conditions of
slow growth~\cite{Cooper1968,Grant2011}. Extending this reasoning to
the case of stationary phase, we can suppose that a higher amount of
genomic DNA per cell would be present in exponential phase with
respect to stationary phase, where we suppose each cell has only one
copy of the genome.
Since typically DNA-binding proteins spend most of their time
non-specifically bound to the genome, genomic DNA acts as a reservoir
of binding sites~\cite{Elf2007}\cite{Bintu2005a}. In other words, the
genome can act as a background reservoir of protein, much as the volume
in a well-stirred system~\cite{Grant2011}. This reasoning leads to
speculate that H-NS would be effectively diluted (with respect to the
genomic background) in exponential phase, compared to stationary
phase, preventing binding to the AT-rich regions of the terminus
macrodomain.

The second possible contribution one can take into consideration is
the effect that the replication forks can have in disrupting the
binding of H-NS to the DNA and the stabilization of its oligomeric
form. One can suppose that H-NS oligomers are disrupted once every
cell cycle when the replication forks pass through a given DNA
stretch. Knowing the number of H-NS molecules per cell (about
20000), and supposing that most are bound nonspecifically to the DNA, we can
estimate the oligomerization speed from a nucleation site from the
diffusion time $\tau = L^2/D$ it takes a nonspecifically bound
molecule to travel the average distance $L$ (which can be estimated
as the mean length of H-NS binding regions in early exponential phase of about $0.5$ $\mu$m).
The diffusion constant $D$ can be estimated as $D=
(4.6\pm1.0)*10^{-14} m^2/s$ ~\cite{diffusion}.  With these figures, we
calculate that the the estimated rate of formation of an H-NS oligomer
after disruption by the replication fork would be of the order of 1
molecule/second. There would be therefore abundant time for
the oligomers to repolymerize after disruption, even in exponential
phase (a 5Kb oligomer would take about ten minutes). This indicates
that H-NS oligomer disruption by replication forks should not be the
dominant process leading the the observed growth phase changes in H-NS
binding.

\section*{Acknowledgments}
We would like to thank L.~Hurst, B.~Bassetti, M.~Osella for useful
discussions, and A.~Seshasayee for feedback and help with the Mock-IP
data. This work was supported by the International Human Frontier
Science Program Organization, grant RGY0069/2009-C.
\footnotesize{
\bibliographystyle{abbrv}
\bibliography{Silent_clusters}
}

\clearpage
\newpage

\section{Graphics and tables}

\newpage

\newpage
\begin{table}[h]
\footnotesize
\centering
\caption{{\bf Coherence of the significant clusters  found from different
    data sets.} Summary of the clusters found analysing
  different data sets. The listed intervals were defined
 by merging the clusters found using different gene sets with a
  window size of $\approx$ 36 kb (128 bins). If a cluster is found in
  a specific dataset it is labeled  by a tick mark.  Clusters that are
  not found using 128 bins, but appear using 
  64 or 256 bins (with distance $\leq0.004$ Mb to the closest cluster found in 128 bin), are labeled by stars. The clusters that are significantly
  close to macrodomain boundaries (distance $\leq0.15$ Mb ) are
  highlighted in red. The clusters that are not close to the
  macrodomain boundaries are found using the lists of transferred genes and genes located in the H-NS binding regions in stationary growth phase, with the exception of the  cluster  located between  0.23 Mb   and 0.35 Mb. }
\begin{tabular}{p{1.7 cm} l p{2 cm} l p{1 cm}}
\hline
Clusters coordinates (Mb) & HGT& H-NS-binding (EE phase)&tsEPODs&Pseudo-genes\\ \hline
0.233-0.348 & \checkmark&\checkmark&*&\checkmark\\ \hline
0.537-0.609 &\textcolor{red}\checkmark&\textcolor{red}\checkmark&{\textcolor{red} \checkmark}&{\textcolor{red}\checkmark}\\ \hline
1.068-1.121 & {\textcolor{red}\checkmark}&{\textcolor{red}-}&{\textcolor{red}-}&{\textcolor{red}-}\\ \hline
1.173-1.252 &{\textcolor{red}\checkmark}&{\textcolor{red} \checkmark}&{\textcolor{red} \checkmark}&{\textcolor{red}*}\\ \hline
1.387-1.448 & \checkmark&-&-&-\\ \hline
1.553-1.679 & \checkmark&-&-&-\\ \hline
2.041-2.087 &{\textcolor{red}*}&{\textcolor{red} -}&{\textcolor{red}-}&{\textcolor{red}\checkmark}\\ \hline
2.444-2.510 &\checkmark&-&-&- \\ \hline
2.747-2.807 &{\textcolor{red}\checkmark}&{\textcolor{red}\checkmark}&{\textcolor{red}\checkmark}&{\textcolor{red} \checkmark}\\ \hline
2.956-3.024 &{\textcolor{red}\checkmark}&{\textcolor{red}\checkmark}&{\textcolor{red} \checkmark}&{\textcolor{red} -}\\ \hline
3.619-3.685 &{\textcolor{red} -}&{\textcolor{red}\checkmark}&{\textcolor{red}\checkmark}&{\textcolor{red} -}\\ \hline
3.765-3.831 &{\textcolor{red}*}&{\textcolor{red}-}&{\textcolor{red}\checkmark}&{\textcolor{red}-}\\ \hline
4.284-4.349 &\checkmark&-&-&-\\ \hline
4.471-4.541 &{\textcolor{red}\checkmark}&{\textcolor{red}*}&{\textcolor{red}-}&{\textcolor{red}\checkmark}\\ \hline
\end{tabular}
\label{result}
\end {table}

\begin{table}[h]\footnotesize
\centering
\caption{{\bf Several of the significant clusters found 
    are close to  macrodomain boundaries.} Comparison of the
  position of clusters found analysing different gene sets with
  macrodomain boundaries and chromosomal sectors. The second and third
  column report the shortest
  distance between one edge of a cluster and the closest macrodomain (dis-MD)
  or chromosomal sector (dis-sector) boundary expressed in Mb. For a
  cluster which overlaps with a boundary, the distance has been
  considered zero.  The fourth and fifth 
  column report the same
  distances normalized by 
  the length of the macrodomain or sector where the cluster is
  found. The normalized distances are indicated by  
  stars. The positions of the
  clusters often lie close to a macrodomain boundary. Here,
  dis-MD$\leq 0.15 $ Mb 
  and dis-MD*$\leq 0.15 $
  are colored in red.   }
\begin{tabular}{p{1.5 cm} p{1.1cm} p{1.1cm} p{1.1cm} p{1.1cm}}
\hline{}
Clusters coordinates (Mb)& dis-MD (Mb) &dis-Sector (Mb)& dis-MD*&dis-Sector*   \\ \hline
0.233-0.348 & 0.183&0.242&0.333&0.148 \\ \hline
0.537-0.609 &{\textcolor{red} {0.000}} &{0.000}&{\textcolor{red}{0.000}}&0.000\\ \hline
1.068-1.121 &{\textcolor{red}{0.089}}&0.208&{\textcolor{red}{0.146}}&0.320\\ \hline
1.173-1.252 &{\textcolor{red}{0.000}}&0.077&{\textcolor{red}{0.000}}&0.118\\ \hline
1.387-1.448 &0.177&0.000&0.182&0.000 \\ \hline
1.553-1.679 &0.343&0.000&0.354&0.000\\ \hline
2.041-2.087 &{\textcolor{red}{0.093}}&0.008&{\textcolor{red}{0.096}}&0.063\\ \hline
2.444-2.510 &0.264&0.173&0.377&0.292\\ \hline
2.747-2.807 &{\textcolor{red}{0.073}}&0.057&{\textcolor{red}{0.104}}&0.096\\ \hline
2.956-3.024 &{\textcolor{red}{0.076}}&0.000&{\textcolor{red}{0.086}}&0.000\\ \hline
3.619-3.685 &{\textcolor{red}{0.075}}&0.024&{\textcolor{red}{0.085}}&0.015\\ \hline
3.765-3.831 &{\textcolor{red}{0.005}}&0.170&{\textcolor{red}{0.005}}&0.104\\ \hline
4.284-4.349 & 0.341&0.689&0.334&0.422\\ \hline
4.471-4.541 & {\textcolor{red}{0.149}}&0.688&{\textcolor{red}{0.146}}&0.421 \\ \hline
\end{tabular}
\label{regions}
\end {table}

\begin{table}[h]
\footnotesize
\centering
\caption{ {\bf Enrichment analysis for functional categories of genes
    located in the clusters reported in Table~\ref{result}.}  List of significant MultiFun gene
  classes for functional genes located inside of the
  clusters. According to the MultiFun data set, 
  3382 genes in the genome (out of 4667) and 757 (out of 1075) within
  the clusters have at least one functional annotation. The first
  column of the table shows the MultiFun category number. The second 
  column shows the number of genes in the MultiFun category (M). The
  third and fourth columns show 
  the number of genes and operons (indicated by $K_1$ and $ K_2$
  respectively) with that MultiFun annotation found in 
  the clusters. 
  The fifth column shows the P-value of a hypergeometric test with
  parameters $K_1$, M, 3382, 757. The last column shows the term 
  related to the MultiFun class. The reported  results are filtered for
  $K_2\geq 3$ and P-values$<0.01$. } 
\begin{tabular}{ l l l l l p{2 cm}}
\hline
Class&M&$K_{1}$&$K_{2}$&P-value&Related term\\ \hline
1.3.7&155&17&8&0.000087& Anaerobic respiration\\ \hline
1.4.1&42&3&3&0.006369&Electron donor\\ \hline
1.5.1.4&9&7&4& 0.000600&Proline\\ \hline
1.6.3.2&17&13&4&0.000003&Lipopolysaccharide synthesis ~~(Core region)\\ \hline
1.6.13&7&5&3&0.007055&Fimbria, pili, curli\\ \hline
1.7.18&4&4&3&0.002495&Betaine biosynthesis\\ \hline
1.8.1&32&15&4&0.001292&Phosphorous metabolism\\ \hline
2.2.3&55&5&5&0.005900&RNA modification\\ \hline
2.3.2&101&5&4&0.000001&Translation\\ \hline
2.3.3&70&8&8&0.008559&Posttranslational modification \\ \hline
2.3.4&91&11&9&0.004914&Protein related (folding)\\ \hline
2.3.8&57&4&3&0.001371&Ribosomal proteins\\ \hline
5.3&59&3&3&0.000230& Motility \\ \hline
5.5.4&14&11&8&0.000011&PH response\\ \hline
6.1 &851&170&142&0.005664&Membrane	\\ \hline
6.4&44&3&3&0.004404&Flagellum\\ \hline
6.5&43&22&13&0.000023&Pilus synthesis\\ \hline
6.6&95&6&5&0.000014& Ribosome\\ \hline
7.1&989&133&109&0.000001&Products location is cytoplasm	\\ \hline
7.2&137&18&16&0.001975&Products location is periplasmic space\\ \hline
7.3&560&107&85&0.005484&Products location is inner membrane\\ \hline
8.1&289&252&141&0.000001&Phage related functions\\ \hline
8.3&66&30&22&0.000016&Transposon related\\ \hline
\end{tabular}
\label{ontology1}
\end {table}

\end{document}


\maketitle

\section*{ Description of Supplementary Figures}

\subsection*{Mock-IP control}

\begin{figure}[htbp]
    \includegraphics[width=120mm]{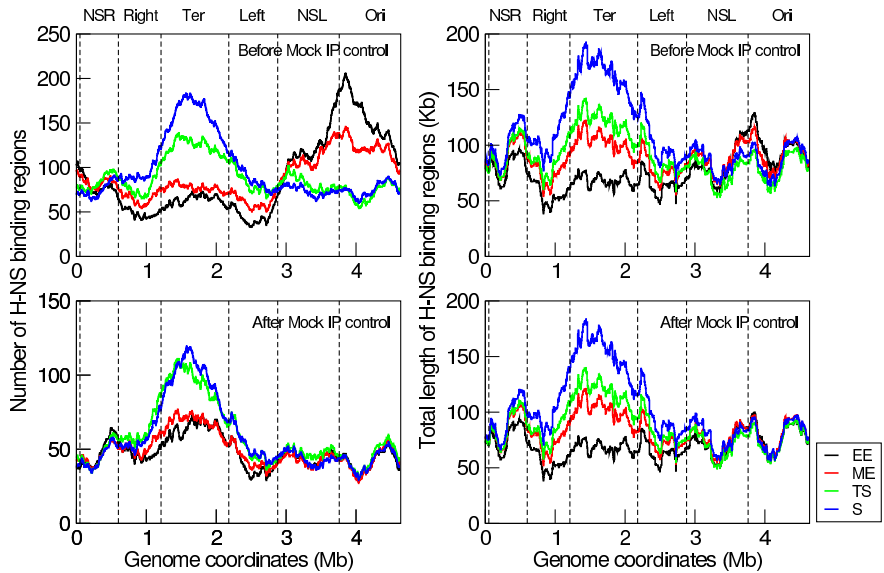}
    \caption{\label{mockip}{\bf The observed growth of H-NS binding
        regions in the Ter region in stationary growth phase is
        unaffected by the Mock-IP control.}  Total
      length (right panels) and number of H-NS binding regions (left panels) along the genome, before and 
      after Mock-IP control in early-exponential (EE),
      mid-exponential (ME), transition to stationary (TS) and
      stationary (S) phases of growth. Because of the gene dosage
      effect in the exponential phase, many spurious binding regions
      around Ori appear, while the Ter region is weakly affected.  Additionally, the
      data for the stationary and transition to stationary phase are
      weakly affected by the Mock-IP control along the whole genome. This indicates that the observed average
      growth of H-NS binding regions around Ter is not a consequence
      of the Mock-IP control.}
 \end{figure}

We analyzed the large-scale organization of H-NS binding regions data
from Kahramanoglou {\it et al}~\cite{aswin} as a function of growth
phase. These binding regions were obtained by comparing the number of
reads mapped to each region, normalized by the total number of reads
obtained for that sample, with the corresponding value from the
Mock-IP experiment using a binomial test~\cite{aswin}.  The Mock-IP
was available only for the mid-exponential phase sample, where the gene dosage
effect is highest.  Since in stationary phase the gene dosage
effect is small or absent, the applied filter might create a bias at
large scales towards the stationary phase dataset. In order to control for
this, we compared the results from binding regions found before
Mock-IP control and after the control. Supplementary Figure~\ref {mockip} shows
the sliding window sums of the length and number of H-NS binding regions along the
genome for binding regions obtained before and after Mock-IP control
in the early-exponential (EE), mid-exponential (ME), transition to
stationary (TS) and stationary (S) phases of growth (window size=500
Kb). The total length is roughly unaffected by
the Mock-IP control, except for minor changes around Ori. The change around
 Ori is larger for the number of binding regions, 
due to the presence of a large number of short spurious regions in the exponential and early exponential phase data.    
However, the results in a wide region around
Ter are robust to the Mock-IP filter for all data sets, for both the number and the total length of bound regions, indicating
that the observed increase of total length of polymerized H-NS is not
a spurious result of the Mock-IP filter.

\subsection* {Growth-phase dependent H-NS binding regions}
\begin{figure}[htbp]
    \includegraphics[width=110mm]{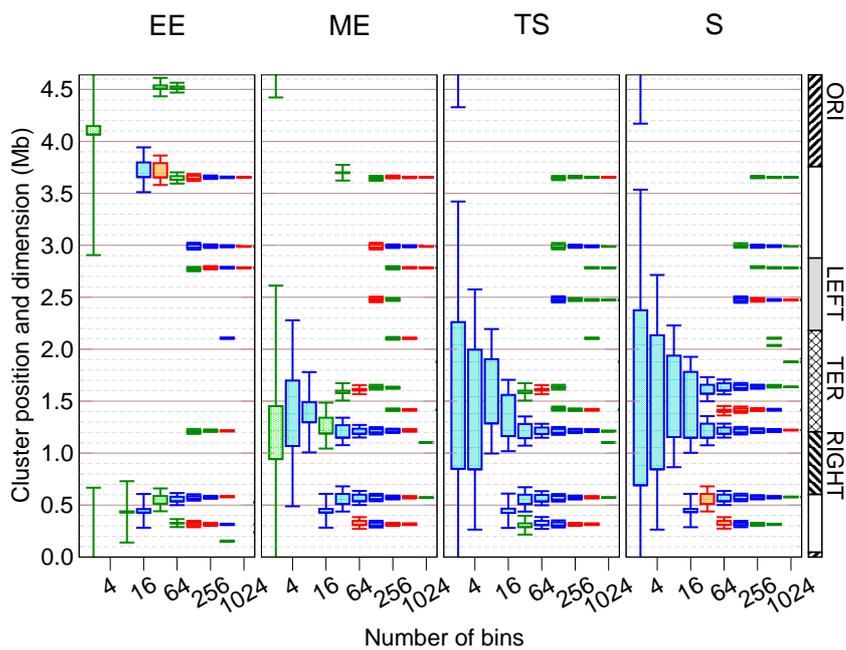}
    \caption{\label{growth}{\bf The number of clusters for genes
        located in the binding regions increase while cells progress
        from early-exponential to stationary growth phase.} Cluster diagrams of genes located in the H-NS
      binding regions in the early-exponential (EE), mid-exponential
      (ME), transition to stationary (TS) and stationary (S) phases of
      growth. The clusters close to the boundaries of the Ter
      macrodomain are preserved during different growth phases, while
      new clusters emerge as the cells progress from early-exponential
      to the stationary phase. }
 \end{figure}

Supplementary Figure~\ref{growth} shows the result of the linear
aggregation analysis for genes located in the 
H-NS binding regions from early exponential to stationary phase. One
can see that the genes located in the H-NS binding regions in early
exponential phase are clustered close to the macrodomain boundaries
and that this cluster pattern is preserved as the cells go from early
exponential to stationary phase. In stationary phase however, new clusters
appear inside the Ter macrodomain.

\subsection* {H-NS binding regions}

\begin{figure}
    \centering
    \includegraphics[width=3.0in]{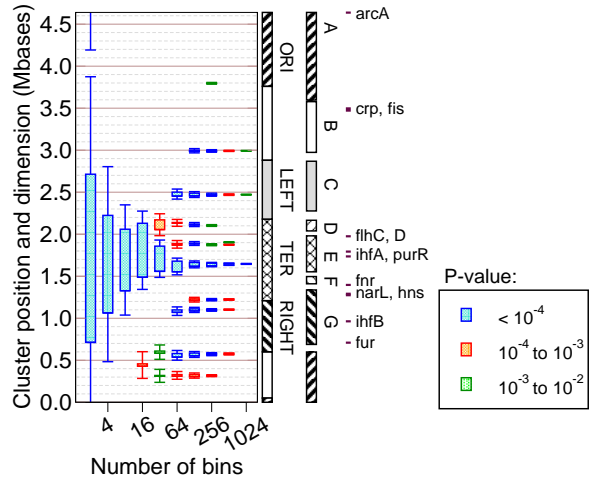}
    \caption{{\bf The H-NS binding regions found by Oshima \emph{et al.}~\cite{oshima} are
      clustered along the genome.} Diagram of the
      statistically significant H-NS bound genes clusters. The plot
      shows that there are significant clusters close to the
      boundaries of macrodomains or Mathelier sectors. The position of
      the clusters correlate with the clusters shown in  Figure 2 of
      the main text.
    }
    \label{hns}
\end{figure}

We performed the clustering analysis for the list of genes associated
to regions of H-NS binding obtained by Oshima {\it et al}~\cite{oshima} using a
high-density oligonucleotide chip (ChIP-chip analysis). The experiment
was performed with the W3110 strain of E.~coli K12 grown in LB
medium. The H-NS genome-wide binding was assessed on exponentially
growing cells. The W3110 strain is very similar to the MG1655 strain
and in their data analysis the genome coordinates of MG1655 were
used. The analysis shows similar clusters to the results presented in
the main text (Supplementary Figure~\ref{hns}).

\subsection*{Genes overlapping with heEPODs 
are clustered along the genome.} 
\begin{figure}
    \centering
    \includegraphics[width=3.0in]{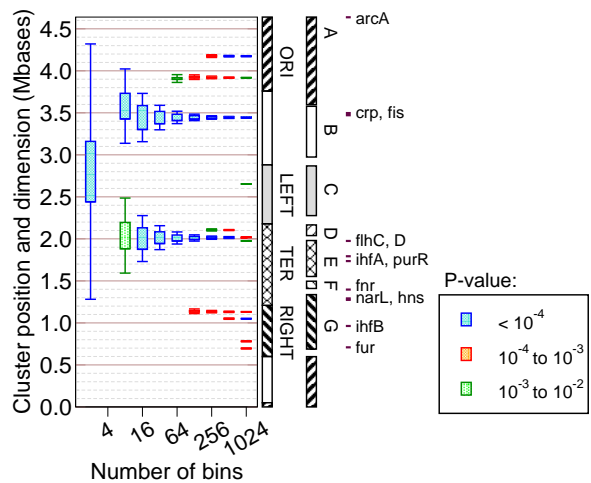}
    \caption{{\bf Genes overlapping with heEPODs are clustered along the
      genome. } Diagram of the statistically
      significant clusters.  The plot shows that there are significant
      clusters of genes overlapping with heEPODs along the
      genome. Three of these clusters include flagella and ribosomal
      genes.  }
    \label{hepod}
\end{figure}

Supplementary Figure~\ref{hepod} shows that the genes overlapping with
heEPODs are also clustered at different observation scales along the
E.~coli genome. The clusters of genes overlapping with heEPODs include
ribosomal and flagella genes, which are highly transcribed, as shown
by Supplementary Table~\ref{ontology1}. Comparison with Figure 4 in
the main text suggests that the clusters of highly transcribed EPODs
are located at the larger distances from macrodomain boundaries with
respect to the tsEPODs clusters.

\subsection*{Most of the clusters found have a lower mean expression level
  than the genomic average.} 
\begin{figure}[htbp]
    \includegraphics[width=110mm]{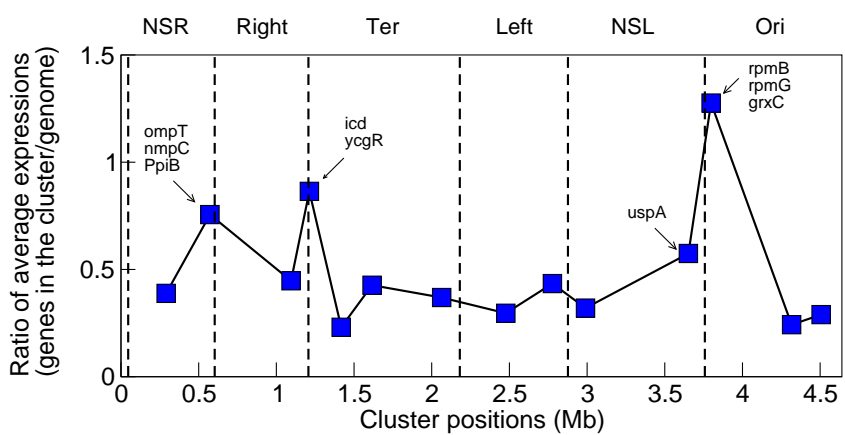}
    \caption{\label{expression}{\bf Most of the clusters found
        have a lower average expression level compared to the average for the E.~coli
        transcriptome.} Ratio of the average
      expression level for each cluster and the average expression
      level of all genes along the genome.  The estimated transcript copy
      number of genes in wild-type E.~coli were obtained from the ASAP
      data base~\cite{asap}, and refers to MG1655 cultured in MOPS minimal
      medium with glucose at 37 degrees to log phase (OD600=0.2).  The
      genes in the clusters that are highly expressed are labeled in
      the figure. Most of the clusters have a significantly lower mean
      expression level compared to the genomic average.  }
 \end{figure}

Supplementary Figure~\ref{expression} shows the ratio of the average expression
level for each cluster and the average expression level of all genes
along the genome. Microarray data sets are extracted from the ASAP
database~\cite{asap} at https:$//$asap.ahabs.wisc.edu. The data set
used for this analysis 
has the transcript copy number of 4220 genes in wild-type E.~coli. The
strain MG1655 was cultured in MOPS minimal with glucose at 37 degrees
to log phase (OD600=0.2).  Most of the clusters are expressed at a
significantly lower level than the average of E.~coli transcriptome.  There
are some genes that are highly expressed and affect the average
expression level for the clusters. For example, ompt and nmpC are
outer membrane related genes which are highly expressed in the data
set that we used. On the other hand, nmpC is reported to be silent in
E.~coli K-12~\cite{nmpc}. rpmB and rpmG are ribosomal proteins, which
are highly transcribed. rpmB and uspA are associated to adaptation to
stress.

\section*{Description of Supplementary Tables}

\subsection* {Coordinates of the macrodomains and chromosomal sectors }
Supplementary Tables~\ref{macrodomain} and \ref{sector} show the
coordinates of the macrdomains and chromosomal sectors used
in this study.

\subsection* {Summary of the clustering for H-NS binding regions and
  horizontally transferred genes} 
Supplementary Tables~\ref{ee},\ref{ss},\ref{htg} summarize the
clustering results for different data sets.

\subsection* {heEPODs clusters are enriched by flagella and ribosomal genes. }
Supplementary Table~\ref{ontology1} shows the results of a
hypergeometric test for the enrichment of MultiFun functional
categories within the lists of genes located in heEPODs clusters. The
results indicate that, as expected, these clusters are enriched by
highly expressed genes, such as flagella, ribosomal proteins, rRNA and tRNA.

\subsection* {Intersection between data sets}
Supplementary Tables~\ref{hyper} and~\ref{size} show the intersection
between different gene lists used in this analysis. The result of a
hypergeometric test (assessing the statistical significance of the
intersection) can be found in Supplementary Table~\ref{hyper}.  The
overlap of the lists is large, but the lists do not coincide, making
the coincidence of the clusters nontrivial.

\subsection* {Summary of the clustering for pseudo-genes} 
Supplementary Table~\ref{pseudo} summarizes the clustering results for
pseudo-genes.

\subsection*{ Some of the long non-lethal deletions and prophages are
  close to the macrodomain boundaries.}
Supplementary Table~\ref{regions} represents the correlation between
the position of clusters found in our analysis and the position of
long non-lethal deletions or prophages.  Some of the clusters
correlate with the prophages and long non-lethal deletions.

\subsection* {Most of the clusters, that are reported in Table 1, are enriched by Prophage related functions.}
Supplementary Table~\ref{ontology2} shows the results of systematic
hypergeometric testing for enrichment of MultiFun functional
categories within the lists of functional genes located in each
cluster that is reported in Table 1. Considering clusters close to or far from the macrodomain
boundaries separately, we did not see particular differences between
the two groups. Not surprisingly, most of the clusters are enriched by
prophage-related functions. The clusters with coordinates 1-17-1.25 Mb
and 3.62-3.68 Mb show enrichment in the term membrane. The clusters
within coordinates 3.6-4.3 Mb show enrichment in the terms membrane,
surface antigens, lipopolysaccharide synthesis, carbon compounds, and anaerobic
respiration.

\footnotesize{
\bibliographystyle{rsc}
\bibliography{supplementary}
}

\newpage

\newpage

\begin{table}[p]
\caption{{\bf Macrodomain coordinates.} }
\centering
\begin{tabular}{l l l l l l l}
\hline
Ori& NSR&Right&Ter&Left&NSL\\ \hline
3.76-0.05&0.05-0.60&0.60-1.21 &1.21-2.18 &2.18-2.88 &2.88-3.76\\
\hline
\end{tabular}
\label{macrodomain}
\end {table}

\begin{table}[p]
\caption{{\bf Chromosomal sectors.} }
\centering
\begin{tabular}{l l l l l l l}
\hline
A&B&C&D&E&F&G\\ \hline
3.59-0.59&2.97-3.57&2.27-2.86&2.03-2.16&1.54-1.97&1.40-1.49&0.68-1.33\\
\hline
\end{tabular}
\label{sector}
\end {table}

\begin{table}[p]
\caption{{\bf H-NS Binding regions in the early exponential growth phase.} }
\centering
\begin{tabular}{l l}
\hline
Number of clusters (bin-size=128)& 6 \\
Total length of binding regions & 648991  \\ 
Total length of binding regions in the clusters & 343221  \\
Total number of genes in the H-NS binding regions& 441  \\ 
Number of H-NS binding genes in the clusters&112\\ 
Number of annotated genes in the clusters&374\\
\hline
\end{tabular}
\label{ee}
\end {table}

\begin{table}[p]
\caption{{\bf H-NS Binding regions in the stationary growth phase.} }
\centering
\begin{tabular}{l l}
\hline
Number of clusters (bin-size=128)& 7 \\
Total length of binding regions & 944735  \\ 
Total length of binding regions in the clusters & 438379  \\
Total number of genes in H-NS binding regions& 748  \\ 
Number of H-NS binding genes in the clusters&217\\ 
Number of annotated genes in the clusters&510\\
\hline
\end{tabular}
\label{ss}
\end {table}

\begin{table}[p]
\caption{{\bf{\small Horizontally transferred genes found using nucleotide composition.}} }
\centering
\begin{tabular}{l l}
\hline
Number of clusters (bin-size=128)& 11 \\
Total number of HGT& 350  \\ 
Number of HGT in the clusters&201\\ 
Number of annotated genes in the clusters&773\\
\hline
\end{tabular}
\label{htg}
\end {table}

\begin{table}[h]
\small
\centering
\caption{ {\bf Enrichment  analysis for functional categories of genes
    located in the heEPODs clusters.}  List of
  significant MultiFunfunctional categories for  genes found within
  the  heEPODs clusters.  The MultiFun data annotate 3382 genes out
  of 4667, and out of 406 genes located in the heEPODs clusters, 326 are
  annotated. The first column shows the MultiFun class number. The second
  column shows the number of genes in the MultiFun class (M). The
  third/fourth column show 
  the number of genes/operons of MultiFun class found in the heEPODs
  clusters ($K_1/K_2$). 
  The fifth column shows P-values obtained from a test with 
  parameters $K_1$, M, 3382, 757. The last column shows the terms 
  related to the MultiFun class. We reported the results with
  $K_2\geq 3$ and P-values$<0.01$. } 
\begin{tabular}{ l l l l l p{3 cm}}
\hline
Class&M&$K_{1}$&$K_{2}$&P-value&Related term\\ \hline
1.3.7&155&4&3&0.000368&Anaerobic respiration\\ \hline
1.6.3.1&14&10&4&0.000001&O antigen\\ \hline
1.6.12&38&31&9&0.000001&Flagella\\ \hline
1.7.10&14&8&4&0.000011&Sugar nucleotide biosynthesis\\ \hline
2.2.5&91&18&11&0.001319&tRNA\\ \hline
2.2.6&26&11&4&0.000010&rRNA, Stable RNA\\ \hline 
2.3.2&101&47&11&0.000001&Translation\\ \hline
2.3.8&57&33&6&0.000001&Ribosomal proteins\\ \hline
3.1.3.1&10&4&4&0.009770&Translation attenuation and efficiency\\ \hline
4.9.B&63&15&5&0.000483&Putative uncharacterized transport protein\\ \hline
4.S.12&12&6&3&0.000391&amino acid\\ \hline
4.S.160&32&15&4&0.000001&protein\\ \hline
5.3&59&33&11&0.000001&Motility\\ \hline
6.1&851&67&44&0.006859&Membrane\\ \hline
6.3&67&14&8&0.002512&Lipopolysaccharide\\ \hline
6.4&44&35&10&0.000001&Flagellum\\ \hline
6.6&95&46&12&0.000001&Ribosome\\ \hline
7.1&989&144&73&0.000001&Products location is cytoplasm\\ \hline
7.3&560&38&25&0.002379&Products location is inner membrane\\ \hline
10&43&17&5&0.000001&cryptic genes\\ \hline
\end{tabular}
\label{ontology1}
\end {table}

\begin{table}[p]
\caption{{\bf Pseudo-genes.} }
\centering
\begin{tabular}{l l}
\hline
Number of clusters (bin-size=128)& 5 \\
Total number of pseudo-genes&212  \\ 
Number of pseudo-genes in the clusters &68\\ 
Number of annotated genes in the clusters&358\\ 
\hline
\end{tabular}
\label{pseudo}
\end {table}

\begin{table}[p]
\caption{{\bf Intersection between different datasets.} Result
  of a hypergeometric test assessing the statistical significance of
  the intersection between datasets. The P-value represents
  the probability of obtaining an intersection of the given size
  selecting two random gene lists (of the same length of the lists in
  consideration) from the total number of genes in the genome.
}
\centering
\begin{tabular}{ l l l l l}
\hline
& HGT& H-NS binding&tsEPODs&Pseudo-genes\\ \hline
HGT & 0& 5.48797e-98&3.41112e-62&1.87785e-06\\
H-NS binding & 5.48797e-98& 0&1.76438e-189&7.49369e-18\\
tsEPODs & 3.41112e-62&1.76438e-189&0&1.44551e-09\\
Pseudo-genes &1.87785e-06&1.44551e-09&7.49369e-18&0\\
\hline
\end{tabular}
\label{hyper}
\end {table}

\begin{table}[p]
\caption{{\bf Intersection between different datasets.} Number of common genes between two different datasets,
  divided by the size of the smallest one.   The datasets have many
  genes in common but they are not identical. Here, we 
  considered genes located in H-NS binding regions in early
  exponential phase. }
\centering
\begin{tabular}{l l l l l}
\hline
& HGT& H-NS binding&tsEPODs&Pseudo-genes\\ \hline
HGT &350/350 &175/350& 107/241&35/212\\
H-NS binding &175/350 & 441/441& 203/241&62/212 \\
tsEPODs &107/241 & 203/241 &241/241 &33/212\\
Pseudo-genes &35/212 &62/212 &33/212 &212/212\\
\hline
\end{tabular}
\label{size}
\end {table}

\begin{table}[t]
\caption{{\bf Some of the significant clusters found in this study overlap
    with the long non-lethal deletions and with the gene from temperate prophages.} Comparison of the position of clusters found analysing different
  datasets with large non-lethal deletions and prophages. The tick
  marks represent overlap between a cluster and a large deletion or a
  prophage.  The position of the clusters correlate
  with the position of some long non-lethal deletions and
  prophages. The clusters that are close to macrodomain boundaries
  (distance $\leq 0.15$ Mb) are colored in red. }
\centering
\begin{tabular}{ l  l l }
\hline
Clusters coordinates (Mb)& Long deletions& Prophages\\ \hline
0.233-0.348 & \checkmark&\checkmark  \\ \hline
0.537-0.609 &{\textcolor{red} \checkmark}&{\textcolor{red} \checkmark} \\ \hline
1.068-1.121 &{\color{red} \checkmark}&{\textcolor{red} -}\\ \hline
1.173-1.252 &{\textcolor{red}-}&{\textcolor{red} \checkmark}\\ \hline
1.387-1.448 &\checkmark&\checkmark \\ \hline
1.553-1.679 &\checkmark&\checkmark\\ \hline
2.041-2.087 &{\textcolor{red} \checkmark}&{\textcolor{red} \checkmark}\\ \hline
2.444-2.510 &-&\checkmark\\ \hline
2.747-2.807 &{\textcolor{red} \checkmark}&{\textcolor{red} \checkmark}\\ \hline
2.956-3.024 &{\textcolor{red} -} & {\textcolor{red} -}\\ \hline
3.619-3.685 &{\textcolor{red} \checkmark}&{\textcolor{red} -}\\ \hline
3.765-3.831 &{\textcolor{red} \checkmark}&{\textcolor{red} -}\\ \hline
4.284-4.349 &\checkmark&-\\ \hline
4.471-4.541 &{\textcolor{red} \checkmark}&{\textcolor{red} \checkmark} \\ \hline
\end{tabular}
\label{regions}
\end {table}

\begin{table}[h]
\footnotesize
\centering
\caption{ {\bf Enrichment analysis for functional annotations of
    genes located in each cluster that is reported in Table 1.}  Summary of the significant MultiFun gene
  classes for each cluster found. The first
  column shows the coordinates of the clusters. The second column
  shows the numbers of functional genes located in each cluster. The
  third column shows the MultiFun class number and related term. Here,
  we considered the results with P-values$<0.01$. We removed the
  results when the number of operons of MultiFun class found in each
  cluster is less than 3.  } 
\begin{tabular}{p{1.8 cm} p{2 cm} p{6.5cm}}
\hline
Clusters coordinates (Mb)&Functional gene count& Functional classes and terms\\ \hline
0.233-0.348 &77&7.1 (Products location is cytoplasm), 7.3 (Products location is inner membrane), 8.1 (Prophage related functions), 8.3 (Transposon related)\\ \hline
0.537-0.609 &61&7.1 (Products location is cytoplasm), 7.4 (Products location is outer membrane), 8.1 (Prophage related functions)\\ \hline
1.068-1.121 &43&8.1 (Phage related functions)\\ \hline
1.173-1.252 &63&  6.1 ( Membrane), 7.1 (Products location is cytoplasm), 8.1 (Prophage related functions),  \\ \hline
1.387-1.448 &52& 8.1 (Prophage related functions)\\ \hline
1.553-1.679 &100& 5.5.4 (PH response), 7.1 (Products location is cytoplasm), 8.1 (Prophage related functions) \\ \hline
2.041-2.087 &36& 2.2.5  (tRNA), 8.1 (Prophage related functions)	\\ \hline
2.444-2.510 &55&7.1 (Products location is cytoplasm), 8.1 (Prophage related functions)\\ \hline
2.747-2.807 &47&8.1 (Prophage related functions)\\ \hline
2.956-3.024 &36&1.7.1  (Unassigned reversible reactions) \\\hline
3.619-3.685 &35& 3.1.2.2 (Activator), 6.1 (Membrane)\\\hline
3.765-3.831 &53& 1.6.3.2 (Lipopolysaccharide), 2.1.4  (DNA repair), 6.3 (Surface antigens), 7.3 (Products location is inner membrane)\\ \hline
4.284-4.349 &46&1.1.1 (Carbon compounds), 1.3.7 (Anaerobic respiration)\\ \hline
4.471-4.541 &53& 2.1.3 (DNA recombination), 8.1 (Prophage related functions), 8.3 (Transposon related)\\ \hline
\end{tabular}
\label{ontology2}
\end {table}